\documentclass[12pt]{article} 
\usepackage{graphicx,floatflt,amssymb,epsf,rotate} 
\usepackage{textcomp}
\usepackage{epsfig}
\begin{document} 
\begin{flushright} 
HRI-P-10-03-001 \\
\end{flushright} 
 
\vskip 30pt 
 
\begin{center} 
{\Large \bf TeV scale double seesaw in left-right symmetric theories}\\
\vspace{0.5in}
{ {\bf Joydeep Chakrabortty$^{\dagger}$
}  }\\ 
\vspace{0.2in}
   {\sl Harish-Chandra Research Institute,\\
Chhatnag Road, Jhunsi, Allahabad  211 019, India} \\
\normalsize 
\end{center} 

\vspace{0.5in}
\begin{abstract}
We extend the Type I and Type III seesaw mechanisms to generate neutrino masses 
within the left-right symmetric theories where parity is spontaneously broken. 
We construct a next to minimal left-right symmetric model where neutrino masses are 
generated through a variant $double$ seesaw mechanism. In our model at least one of the
triplet fermions and the right handed neutrinos are at TeV scale and others are heavy. 
The phenomenological aspects and testability of the TeV scale particles at collider experiments are discussed. The decays of heavy fermions leading to leptogenesis are pointed out. 

\end{abstract}

\vskip 3 true cm

{\small \footnoterule{$^{\dagger}$ e-mail: joydeep@hri.res.in}}

\pagebreak

\renewcommand{\thesection}{\Roman{section}} 
\setcounter{footnote}{0} 
\renewcommand{\thefootnote}{\arabic{footnote}} 
\noindent

\section{Introduction}
The Large Hadron Collider (LHC) is being operated to test the Standard Model (SM) 
and the theory beyond it. Physics at TeV scale is expected to be probed. Will that be enough 
to get the flavour of the physics beyond the SM? Physics in the neutrino sector 
might be a smoking gun to probe the high scale theories. The small but nonzero neutrino masses 
are unexplained within the SM. There are several models, beyond SM, which attempt to explain the origin of the neutrino masses. Seesaw mechanism where light neutrino masses are generated by integrating out the heavy 
particles is one of the popular and well established models. Studies are going on to explore different possibilities within the seesaw models, namely Type I, II, III, and Inverse seesaw depending on the nature of the $heavy$ particles. These $heavy$ particles carry the signatures of the high scale theories and their need strongly motivates to go beyond the SM. 

In case of Type I seesaw mechanism~\cite{TypeI}, one should have
atleast two fermion singlets $\nu_R$ (right-handed neutrinos) 
and the neutrino masses read as $m_{\nu_L}^I \simeq m_D^2/ M_{\nu_R}$,
where $m_D$ is the Dirac mass term and $M_{\nu_R}$ are the 
Majorana (lepton number violating) masses of the right handed neutrinos. If $m_D$ $\approx$ 100 GeV, and $M_{\nu_R} \approx 10^{13}$ GeV, one obtains the natural value for the neutrino masses $m_\nu \approx 1$ eV.

In Type II seesaw mechanism~\cite{TypeII} the SM is extended by 
an $SU(2)_L$ triplet Higgs $\Delta$. In this scenario the neutrino
masses are given by $m_{\nu_L}^{II} \simeq Y_\nu  v_{\Delta}$, where
$v_{\Delta}$ is the vacuum expectation value (vev) of the neutral component of the triplet 
and $Y_\nu$ is the Yukawa coupling. $v_{\Delta} \simeq \mu v^2/M_{\Delta}^2$, 
where $M_{\Delta}$ is the mass of the triplet and $\mu$ is
the trilinear coupling between the SM Higgs and the triplet. Light neutrino masses $m_\nu \approx 1$ eV 
if we assume $Y_\nu \approx 1$, $v \approx 100$ GeV, and $\mu \sim M_{\Delta} \approx 10^{13-14}$ GeV.

For the Type III seesaw mechanism~\cite{TypeIII} one needs to add at least two extra matter
fields in the adjoint representation of $SU(2)_L$ with zero hypercharge
to generate neutrino masses, $m_{\nu_L}^{III} \simeq M_D^2/M_{\Sigma}$. 
Here $M_{\Sigma}$ stands for the mass of the fermionic triplets and $M_D$ is 
the Dirac coupling. 

We start with left-right symmetric theory and discuss a TeV scale 
$double$ seesaw mechanism for neutrino mass generation. In section II, we propose a model 
where right handed neutrino masses are generated through a Type III seesaw and light neutrino masses 
are the outcome of a Type I + Type III seesaw mechanisms.
In the later sections, we discuss the phenomenological implications of this model. In this context we 
point out the possible decay modes of the triplet fermions and the right handed neutrinos. Leptogenesis, through the decays of right handed neutrinos and triplet fermions, is discussed within the context of this model.

\section{Double seesaw in left-right symmetric model}

The so-called left-right symmetric models among the 
most appealing extensions of the SM where one can understand 
the origin of parity violation in a simple way and we can generate
neutrino masses. The simplest theories are based on the gauge group
$ SU(2)_L \otimes SU(2)_R \otimes SU(3)_C \otimes U(1)_{B-L}$.
Here $B$ and $L$ stand for baryon and lepton number, respectively.
The matter multiplets for quarks and leptons are given by
\begin{equation}
Q_L = \left(
\begin{array} {c}
u_L \\ d_L
\end{array}
\right) \ \equiv \ (2,1,3,1/3),~
Q_R = \left(
\begin{array} {c}
 u_R \\ d_R
\end{array}
\right) \ \equiv \ (1,2,3,1/3),
\end{equation}
\begin{equation}
l_L = \left(
\begin{array} {c}
 \nu_L \\ e_L
\end{array}
\right) \ \equiv \ (2,1,1,-1),~
l_R = \left(
\begin{array} {c}
 \nu_R \\ e_R
\end{array}
\right) \ \equiv \ (1,2,1,-1).
\end{equation}

In our model we have two Higgs doublets:
\begin{equation}
H_L = \left(
\begin{array} {c}
h_{L}^{+} \\ h_{L}^{0}
\end{array}
\right) \ \equiv \ (2,1,1,1),~
H_R = \left(
\begin{array} {c}
 h_{R}^{+} \\ h_{R}^{0}
\end{array}
\right) \ \equiv \ (1,2,1,1),
\end{equation}
and a bidoublet Higgs:
\begin{equation}
\Phi = \left(
\begin{array} {cc}
 \phi_1^0   &  \phi^+_2 \\
 \phi^-_1  & \phi_2^0
\end{array}
\right) \equiv (2, 2, 1, 0), ~~~ and~~~ \tilde{\Phi}= \sigma_2 \Phi^* \sigma_2.
\end{equation}
Under the left-right parity transformation one has the following relations
\begin{equation}
F_L \longleftrightarrow F_{R}
\end{equation}
where $F$ = $Q$, $l$, $H$.

The relevant Yukawa interactions for quarks in this 
context are given by
\begin{eqnarray}
{\cal L}_Y^{quarks} &=& \bar{Q}_L( Y_1 \Phi  +  Y_2 \tilde{\Phi}) Q_R  + h.c.
\end{eqnarray}

Once the bidoublet gets the vev the quark mass matrices read as
\begin{eqnarray}
M_U & = & Y_1 v_1 \ + \ Y_2 v_2^*~~~ \mbox{and} ~~~
M_D =  Y_1 v_2 \ + \ Y_2 v_1^*,
\end{eqnarray}
with $v_1= \langle \phi_1^0 \rangle$ and $v_2 = \langle \phi_2^0 \rangle$.
In the case of the bidoublet one has the following transformation under the left-right parity
\begin{equation}
\Phi \longleftrightarrow \Phi^\dagger
\end{equation}
and $Y_1 = Y_1^\dagger$ and $Y_2 = Y_2^\dagger$.

In this context the charged lepton masses are generated through the interactions
\begin{equation}
{\cal L}_l= \bar{l}_L ( Y_3  \Phi  +  Y_4 \tilde{\Phi}) l_R  + h.c.
\end{equation}
and the relevant mass matrix is given by
\begin{equation}
M_\ell= Y_3  v_2  + Y_4 v_1^*  ~~(\ell=e,\mu,\tau).
\end{equation}
At the same time Dirac mass matrix for the neutrinos is written as:
\begin{equation}
m_\nu^D= Y_3 v_1  +  Y_4 v_2^*.
\end{equation}

However, in this case one has the same situation as in the SM plus
right-handed neutrinos where we can assume a small Dirac Yukawa coupling
for neutrinos. In this model right handed neutrino can not have any tree level mass. 
Once the $H_L$ and $H_R$ acquire the vevs, $v_L$ and $v_R$ respectively, left-right symmetry 
is broken. In our model $v_L$=0 and the vev ($v_R$) of $H_R$ sets the scale where $SU(2)_R\otimes U(1)_{B-L}$ is broken to $U(1)_Y$.

The realization of the variant $double$ seesaw mechanism has not been
studied in the context of left-right symmetric theories. 
In order to realize this mechanism one has to introduce 
fermionic triplets (one for each family):
\begin{equation}
\Sigma_L = \frac{1}{2} \left(
\begin{array} {cc}
 \Sigma_L^0  &  \sqrt{2} \Sigma^+_L \\
 \sqrt{2} \Sigma^-_L  & - \Sigma_L^0
\end{array}
\right) \ \equiv \ (3,1,1,0),
\end{equation}
and
\begin{equation}
\Sigma_R = \frac{1}{2} \left(
\begin{array} {cc}
 \Sigma_R^0  &  \sqrt{2} \Sigma^+_R \\
 \sqrt{2} \Sigma^-_R  & - \Sigma_R^0
\end{array}
\right) \ \equiv \ (1,3,1,0),
\end{equation}
and Higgses in the mixed representations of $SU(2)_L$ and $SU(2)_R$,
respectively.
\begin{equation}
\bigtriangleup_1 \equiv \ (3,2,1,-1) ~ \mbox{and} ~ \bigtriangleup_2 \equiv \ (2,3,1,-1).
\end{equation}
Under left-right parity transformation
one has the following relations
\begin{equation}
\Sigma_L \longleftrightarrow \Sigma_R  ~\mbox{and}~  \bigtriangleup_1 \longleftrightarrow \bigtriangleup_2.
\end{equation}
In this case the relevant interactions (for the $double$ seesaw mass matrix) are given by
\begin{eqnarray}
{\cal L}^{III}_\nu &=& \ Y_5 \left( l_L^T \ C \ i \sigma_2 \ \Sigma_L \ H_L
\ + \ \ l_R^T \ C \ i \sigma_2 \ \Sigma_R \ H_R \right) \nonumber \\
& + & \ M_\Sigma \ Tr \left( \Sigma_L^T \ C \ \Sigma_L \ + \ \Sigma_R^T \ C \Sigma_R \right) +  
 \ Y_6 \ \bar{l}_L  \bigtriangleup_2 \Sigma_R  + \ Y_7 \ \bar{l}_R  \bigtriangleup_1 \Sigma_L + h.c.
\end{eqnarray}

Therefore, once the Higgses $\bigtriangleup_1$ and $\bigtriangleup_2$ get the vevs, $V_1$ and $V_2$ 
respectively, $SU(2)_L \otimes SU(2)_R$ is broken spontaneously. In the case when $V_1 \sim 0$ 
{\footnote{As the vev affects $\rho$ parameter, we assign a very small value to it. 
Because of lack of the couplings with other neutral particles ($\nu_R, \Sigma_R^0$), 
$\Sigma_L^0$ will only have negligible impact in the light neutrino masses.}} 
and $V_2 \neq 0$ one finds that the mass matrix in the basis 
$\left(\nu_L, \ \nu_R, \ \Sigma^0_R \right)$ reads as:
\begin{equation}
M_{\nu}^{III} = \left(
\begin{array} {ccc}
 0 & m_\nu^D  & Y_6 V_2
\\
(m_\nu^D)^T & 0 & Y_5 v_R
\\
Y_6^T V_2 & Y_5^T v_R & M_{\Sigma}
\end{array}
\right).
\end{equation}
As one expects the neutrino masses 
are generated through the Type I + Type III seesaw mechanisms
and one has a $double$ seesaw mechanism since the mass of
the right-handed neutrinos are generated through
the Type III seesaw once we integrate out $\Sigma_R^0$.

Assuming that $M_\Sigma >> Y_5 v_R  >> m_\nu^D, Y_6 V_2$ one gets
\begin{equation}
m_{\nu_L} = \frac{1}{v_R^2(Y_5^T Y_5)}[m_{\nu}^D M_{\Sigma} (m_{\nu}^D)^T-v_R V_2 m_{\nu}^D (Y_6Y_5)^T-v_RV_2 (Y_6Y_5)(m_{\nu}^D)^T]
\end{equation}
with right handed neutrino masses
\begin{equation}
M_{\nu_R} = v_R^2  Y_5  \left( M_{\Sigma} \right)^{-1}  Y_5^T.
\end{equation}
Notice the double-seesaw mechanism, where the mass of the right-handed neutrinos
are generated once the fermionic triplet is integrated out (Type III seesaw),
and later the light neutrinos get the mass through the usual Type I mechanism.
Here we stick to the case $v_L=0$ since it has been shown
in~\cite{GoranLR} that this solution corresponds to the
minimum of the scalar potential. Here $V_1$ does acquire a very feeble vev, so the $\rho$ 
parameter in the SM will be negligibly affected. 
Eqs. (10) and (11) allow us to consider $m_{\nu}^D$ to be of the same order as the charged 
lepton masses. This helps to fix the scales, $M_{\Sigma}$ and $v_R$, 
compatible with the light neutrino masses. The first term in the expression for light neutrino masses 
reads as:
\begin{eqnarray}
&&m_{\nu_e}^D \sim M_e:\;\;\;\;m_{\nu_1} = M_e^2/M_{\nu_{R_1}}\;,\;\;\mbox{with}\;\; m_{\nu_1} =0.1~\mbox{eV}, 
M_{\nu_{R_1}} \simeq 2.5~\mbox{TeV}\nonumber\\
&&m_{\nu_{\mu}}^D \sim M_\mu:\;\;\;\;m_{\nu_2} = M_\mu^2/M_{\nu_{R_2}}\;,\;\;\mbox{with}\;\; m_{\nu_2} =0.1~\mbox{eV}, M_{\nu_{R_2}} \simeq 10^8 ~\mbox{GeV}\\
&&m_{\nu_{\tau}}^D \sim M_\tau:\;\;\;\;m_{\nu_3} = M_\tau^2/M_{\nu_{R_3}}\;,\;\;\mbox{with}\;\; m_{\nu_3} =0.1~\mbox{eV}, M_{\nu_{R_3}} \simeq  10^{10}~\mbox{GeV}\nonumber.
\end{eqnarray}
Second and third terms can be expressed as:
\begin{equation}
m_{\nu_\ell}\cong\frac{Y_6 V_2 M_\ell}{Y_5 v_R} ~~~~~~(\ell =e,\mu, \tau).
\end{equation}

In our model we consider $Y_6,Y_5 < 1$, $v_R \sim 3 \times 10^{6}$ GeV in Eq. (19)
and these lead to triplet fermion masses : $M_{\Sigma_1}\sim 4 \times 10^9$ GeV, 
$M_{\Sigma_2}\sim 10^5$ GeV, and $M_{\Sigma_3}\sim 500$ GeV. $V_2 \sim 1-10$ MeV in Eq. (21) 
is compatible with the light neutrino masses. In the next section, to test this model, 
we explore the accessible decay modes of TeV scale triplet fermions 
and perform a brief comparative study among them. Then we calculate the CP asymmetries 
leading to leptogenesis from the decays of the heavy fermions.


\section{Signals of Triplet fermions}

In this model we have TeV scale triplet fermions - neutral Majorana particle $\Sigma^0$ and 
charged particles $\Sigma^\pm$. They have gauge interactions which favour to be studied at 
LHC than the singlet fermion models. These TeV scale triplet fermions can be produced at LHC 
\cite{Bajc,Strumia2, deAguila}. The charged and the neutral fermions are produced through the processes:
\begin{eqnarray}
&& pp \rightarrow q\bar{q} \rightarrow Z^*/\gamma^* \rightarrow\Sigma^+ \Sigma^- \nonumber \\
&& pp \rightarrow q\bar{q'}  \rightarrow W^* \rightarrow \Sigma^\pm \Sigma^0. \nonumber
\end{eqnarray}
$\Sigma^0$, having $T_3$=Y=0, pair can not be produced as they do not couple to $Z$ boson. 
The production crosssection depend on the masses of the triplet fermions. The charged 
and the neutral components of the triplet fermions can be treated almost degenerate 
inspite of having small splitting for their phenomenological aspects. 

One needs to study the decays of the triplet leptons to specify detection signals.
The decay modes for the heavy triplet leptons are mainly
\begin{eqnarray}
&&\Sigma^0 \to \;\;\;\;l^\pm W^\mp\;,\;\;\;\;\nu Z\;,\;\;\;\;\nu h\;,\nonumber\\
&&\Sigma^\pm \to\;\;\;\; l^\pm Z,\;\;\;\;\nu W^\pm\;,\;\;\;\;l^\pm h\;.\nonumber
\end{eqnarray}
The $\Sigma_0$ and $\Sigma^\pm$ decay widths are given in~\cite{Bajc, Strumia2, deAguila}:
\begin{eqnarray}
\Gamma(\Sigma^\pm \to \ell^\pm h ) = \Gamma(\Sigma^0 \to \nu_l h ) & =& \frac{g^2}{64\pi} |V_{l\Sigma}|^2 \frac{M_{\Sigma}^3}{M_W^2}
(1- \frac{m_h^2}{M_\Sigma^2})^{2} ,\nonumber \\
\Gamma(\Sigma^\pm \to \ell^\pm Z )= \Gamma(\Sigma^0 \to \nu_l Z ) &=& \frac{g^2}{64\pi c_W^2} |V_{l\Sigma}|^2 \frac{M_{\Sigma}^3}{M_Z^2}
(1- \frac{m_Z^2}{M_\Sigma^2}) (1+\frac{m_Z^2}{M_\Sigma^2}-2\frac{m_Z^4}{M_\Sigma^4}) ,\nonumber \\
\Gamma(\Sigma^\pm \to  \bar{\nu}_\ell W^\pm ) =2~\Gamma(\Sigma^0 \to  \ell^\mp W^\pm ) &=& \frac{g^2}{32\pi} |V_{l\Sigma}|^2 \frac{M_{\Sigma}^3}{M_W^2}
(1- \frac{m_W^2}{M_\Sigma^2}) (1+\frac{m_W^2}{M_\Sigma^2}-2\frac{m_W^4}{M_\Sigma^4}) ,\nonumber
\end{eqnarray}
where $V_{l\Sigma}$ is the small mixing $\sim$ $10^{-6}-10^{-8}$.
Here we list the possible final states and their respective branching ratios (BR) \cite{deAguila}:

i) {\bf 6 charged leptons $l^\pm l^\pm l^\pm l^\mp l^\mp l^\mp$}:\\
$\Sigma^+ \Sigma^-  \to \ell^+ Z \, \ell^- Z$ with $Z \to \ell^+ \ell^-$ 
(BR = $3.5 \times 10^{-4}$). 
This has no SM background but the cross section is very small. To explore this channel one
needs very large integrated luminosities $\sim ~300 ~ fb^{-1}$

ii) {\bf 5 charged leptons $l^\pm l^\pm l^\mp l^\mp l^+$}:\\
$\Sigma^+ \Sigma^0 \to \ell^+ Z \, \ell^\pm W^\mp$ with $Z \to \ell^+ \ell^-, W \to \ell \nu$ 
 (BR = $2.2 \times 10^{-3}$);\\
$\Sigma^+ \Sigma^0 \to \ell^+ Z\, \nu Z $ with $Z \to \ell^+ \ell^-$  (BR = $3.5 \times 10^{-4}$).
This has also very small cross section.

iii) {\bf 4 charged leptons + $jets$}:\\
We can have two possibilities: (a) three same sign leptons and one opposite sign lepton ($l^\pm l^\pm l^\pm l^\mp$+ $jets$) for the processes like,
$\Sigma^+ \Sigma^0 \to \ell^+ Z \, \ell^+ W^-$ with $Z \to \ell^+ \ell^-$, $W \to q \bar q'$ 
  (BR = $3.4 \times 10^{-3}$), and  
(b) two pairs of leptons having mutually opposite signs ($l^+ l^+ l^- l^-$+$jets$/MET) for the processes like,
$\Sigma^+ \Sigma^- \to \ell^+ Z \, \ell^- Z $ with $ZZ \to \ell^+ \ell^- q \bar q / \nu \bar\nu$
 (BR = $9.1 \times 10^{-3}$);  
$\Sigma^+ \Sigma^- \to \ell^+ Z \, \ell^- H / \ell^+ H \, \ell^- Z $ with $Z \to \ell^+ \ell^- , H \to q \bar q$ (BR = $6.9 \times 10^{-3}$);  
$\Sigma^+ \Sigma^- \to \nu W^+ \ell^- Z / \ell^+ Z \nu W^- $ with $Z \to \ell^+ \ell^- , W \to \ell \nu$  (BR = $4.5 \times 10^{-3}$);  
$ \Sigma^\pm \Sigma^0 \to \ell^\pm Z \, \ell^- W^+ $  with $ Z \to \ell^+ \ell^- , W \to q \bar q'$
 (BR = $3.4 \times 10^{-3}$).

v) {\bf 3 charged leptons $l^\pm l^\pm l^\mp$}:\\
For the discovery, signals with this final state is less suppressed by the background 
than other cases. $ \Sigma^+ \Sigma^0 \to \ell^+ Z \, \ell^\pm W^\mp $
with $Z \to q \bar q / \nu \bar \nu, W \to \ell
\nu$ (BR = $2.8 \times 10^{-2}$); $ \Sigma^+ \Sigma^0 \to \ell^+ H \, \ell^\pm W^\mp$ with 
$H \to q \bar q , W \to \ell \nu$ (BR = $2.2 \times 10^{-2}$); $\Sigma^+ \Sigma^0 \to \bar \nu W^+ \, \ell^\pm W^\mp$ with $W \to \ell \nu$ (BR = $7.2 \times 10^{-3}$).

vi) {\bf 2 charged leptons + $jets$}:\\
Lepton flavour violation manifests in the processes like: $pp \rightarrow l_1\bar{l}_2ZW^+$ 
and $pp \rightarrow \ell_1\bar{\ell}_2ZZ$, where $l_{1,2}$ are two different lepton flavours. 
The decays of $Z$ into two leptons is the cleanest signal. Here we consider only the hadronic decay modes 
of all the vector bosons leading to  $l^+l^- jjjj$. 
This might be the less clean channel but have largest event rate. We have listed below the final states 
with opposite sign dileptons (OSD): 
 $\Sigma^+ \Sigma^0 \to \ell^+ Z \, \ell^- W^+$ with
  $\to q \bar q / \nu \bar \nu, W \to q \bar
  q'$ (BR = $4.3 \times 10^{-2}$); 
$\Sigma^+ \Sigma^0 \to \ell^+ H \, \ell^- W^+$ with
  $H \to q \bar q , W \to q \bar q'$ (BR = $3.3 \times 10^{-2}$); 
$\Sigma^+ \Sigma^0 \to \bar \nu W^+ \, \ell^- W^+$ with
  $W W\to \ell \nu q \bar q'$ (BR = $4.3 \times 10^{-3}$);  
 $\Sigma^+ \Sigma^- \to \ell^+ Z \, \ell^- Z$ with
  $Z \to q \bar q / \nu \bar \nu$ (BR = $5.9 \times 10^{-2}$); 
 $ \Sigma^+ \Sigma^- \to \ell^+ Z \, \ell^- H / \ell^+ H \, \ell^- Z$ with
  $Z \to q \bar q / \nu \bar \nu , H \to q \bar
  q$ (BR = $8.9 \times 10^{-2}$); 
$\Sigma^+ \Sigma^- \to \ell^+ H \, \ell^- H$ with
  $H \to q \bar q$ (BR = $3.4 \times 10^{-2}$).\\
We can have also same sign dileptons, $l^\pm l^\pm$, (SSD) in the final states: 
 $\Sigma^+ \Sigma^0 \to \ell^+ Z \, \ell^+ W^-$ with
  $Z \to q \bar q / \nu \bar \nu, W \to q \bar
  q'$ (BR = $4.3 \times 10^{-2}$); 
$\Sigma^+ \Sigma^0 \to \ell^+ H \, \ell^+ W^- $,
 with $H \to q \bar q , W \to q \bar q'$ (BR = $3.3 \times 10^{-2}$); 
$\Sigma^+ \Sigma^0 \to \bar \nu W^+ \, \ell^+ W^-$ with
  $W^+ \to \ell \nu, W^- \to q \bar q'$ (BR = $2.1 \times 10^{-3}$). 
These SSD final state processes are lepton number violating as well as lepton flavour violating. 
This signal is less suppressed by the background as can not be faked by $t\bar{t}$ production.

viii) {\bf 1 charged lepton $\ell^\pm$ + $jets$}:\\ 
This channel have large cross section but suppressed by background. The processes leading to this kind of final states are: 
$\Sigma^+ \Sigma^0 \to \bar \nu W^+ \, \ell^\mp W^\pm $ with
  $W \to  q \bar q'$ (BR = $1.28 \times 10^{-1}$);
$\Sigma^+ \Sigma^0 \to \ell^+ Z/H \, \nu Z/H $ with $Z \to  q \bar q'/\nu \bar \nu,H \to q \bar
  q$ (BR = $1.82 \times 10^{-1}$);
$\Sigma^+ \Sigma^- \to \ell^+ Z/H \, \nu W^-$ with $Z \to q \bar q/\nu \bar \nu,H \to q \bar q,W \to q \bar q'$ (BR = $1.52 \times 10^{-1}$);
$\Sigma^+ \Sigma^- \to \bar \nu W^+ \, \ell^- Z/H $ with $Z \to q \bar q/\nu \bar \nu,H \to q \bar q,W \to q \bar q'$ (BR = $1.52 \times 10^{-1}$).

\section{Signals of the right handed neutrinos}

We have TeV scale right handed neutrino in our model and their signatures can be grabbed at the LHC 
\cite{Han}. The main production channels of the single heavy right handed neutrino are:
\begin{eqnarray}
p p  & \rightarrow & q \bar{q'} \rightarrow W^* \rightarrow \ell^\pm \nu_R, \nonumber \\
    &  \rightarrow & q \bar{q} \rightarrow Z^* \rightarrow \nu \nu_R, \nonumber \\
    &  \rightarrow & gg \rightarrow H^* \rightarrow \nu \nu_R, \nonumber
\end{eqnarray}
and the pair production signal is:
\begin{equation}
 p p \rightarrow q \bar{q} \rightarrow Z^* \rightarrow \nu_R \nu_R. \nonumber
\end{equation}
The production cross sections are very much suppressed by the mixing matrix elements ($V_{l\nu}$) 
and the SM backgrounds. Here we briefly discuss the possible decay modes:
\begin{equation}
\nu_{R_{i}} \rightarrow l^{\pm} l^{\pm} W^{\mp}W^{\mp} ~~~(l=e, \mu, \tau, i=1,2,3). 
\end{equation}
This is a lepton number violating process. The decay widths for $\nu_R$ decays are same as for 
the $\Sigma^0$ in the above section.
If we consider that $W'$s decay hadronically then in the final state we will have same sign dileptons 
($SSD$) and 4 $jets$. There could be other comparatively clean channels like $l^{\pm} l^{\mp}+ ~4~ jets$, but these are not lepton number violating processes. 
We can also look at the leptonic decay of one of the $W'$s 
\begin{equation}
\nu_{R_{i}} \nu_{R_{j}} \rightarrow l^{\pm} l^{\pm} W^{\mp}W^{\mp}\rightarrow l^{\pm} 
l^{\pm}l^{\mp}~+~ MET+~ 2 ~jets. 
\end{equation}
This might be a faithful decay mode to reconstruct the mass of the right handed neutrino 
as there is a single source of missing transverse energy (MET). The combinatorial background 
can be reduced by minimizing the difference of the masses of two right handed neutrinos.

\section{Leptogenesis}
In present universe the baryon number density is much larger than expectation- baryon asymmetry problem. 
Leptogenesis \cite{leptogenesis} is a good candidate for producing a lepton asymmetry which can be 
transfered to a baryon asymmetry through a sphaleron effect. 
In order to have leptogenesis the lepton number asymmetry ($\bigtriangleup L\neq$0) must be generated and persist the evolution of the universe to be converted into baryon asymmetry. CP violating interactions generate lepton number asymmetry when a heavy Majorana fermion decays.
One important ingredient that determines the baryon asymmetry produced by thermal leptogenesis is the
CP asymmetry $\varepsilon_i$. In this model CP asymmetry can be produced from decays of  heavy 
right handed neutrinos{\footnote {In this section we define right handed neutrinos as $N_i$ and $H$ is contained in $\Phi$.}} and triplet fermions. 
It is useful to decompose the CP asymmetry in $N_i$ decays 
as the sum of a Vertex contribution ($V_j$) and of a Self-energy contribution ($S_j$) 
\cite{Pilaftsis, Hambye1, Hambye2}. 
\begin{equation}\label{eps}
\varepsilon_{N_i} \equiv \frac{\Gamma({N_i} \rightarrow l H^*) -
\Gamma( N_i \rightarrow \bar{l} H)}
{\Gamma({N_i} \rightarrow l H^*) + \Gamma( N_i \rightarrow \bar{l} H)}=
-\sum_{j}\frac{3}{2}
  \frac{M_{N_i} }{M_{N_j} }\frac{\Gamma_{N_j} }{M_{N_j} }
  I_j\frac{2 S_j + V_j}{3} 
\end{equation}
where
\begin{equation}\label{IGamma}
I_j = \frac{ \hbox{Im}\,[ (Y_{N} Y_{N}  ^\dagger)_{ij}^2 ]}
{|Y_{N} Y_{N} ^\dagger |_{ii} |Y_{N} Y_{N} ^\dagger |_{jj}}
 \, ,\qquad
\frac{\Gamma_{N_j}}{M_{N_j}} = \frac{|Y_{N} Y_{N} ^\dagger |_{jj}}{8\pi}
\equiv \frac{\tilde{m}_j M_{N_j}}{8\pi v^2}
\,,
\end{equation}
and where
\begin{equation}
S_j = \frac{M^2_{N_j}  \Delta M^2_{ij}}{(\Delta M^2_{ij})^2+M_{N_i} ^2
   \Gamma_{N_j} ^2} \,, \,\,\,\,\,
V_j = 2 \frac{M^2_{N_j} }{M^2_{N_i}}
\bigg[ \big(1+\frac{M^2_{N_j} }{M^2_{N_i}}\big)\log\big(1+
\frac{M^2_{N_i}}{M^2_{N_j} }\big)
- 1 \bigg],
\end{equation}
with $\Delta M^2_{ij}=M^2_{N_j}-M^2_{N_i}$ and $Y_N$ is the Yukawa coupling for the 
right handed neutrinos (see eq. 9).
The factors $S_j$ ($V_j$) comes from the one-loop 
self-energy (vertex) contribution to the decay widths, Fig.~1. 
The $I_j$ factors are the CP violating coupling combinations entering 
in the asymmetry. The CP asymmetry in hierarchical limit ($M_{N_2} << M_{N_{3}}$) is given as:
\begin{equation}
\varepsilon_{N_2}=-\frac{3}{16 \pi} \frac{M_{N_2}}{M_{N_3}}
\frac {Im[(Y_{N} Y_{N}^{\dagger})_{23}^2]}
{\sum_i |(Y_{N})_{2i}|^2} \,.
\label{nu-nu}
\end{equation}

\begin{figure}[htb]
\centering \vspace*{3mm} \hspace*{1mm}
\epsfig{figure=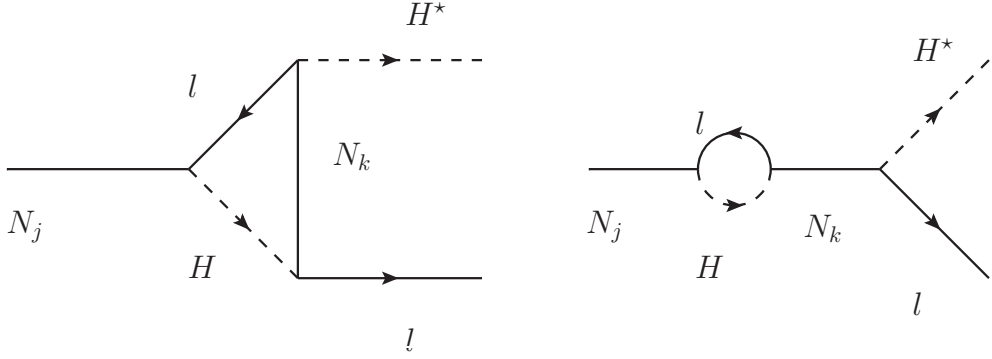,width =0.8\columnwidth} \vspace*{-1mm}
\caption{One loop vertex and self-energy diagrams contributing to
the asymmetry via decay of RH neutrinos, $M_{N_j} << M_{N_{k}}$.}
\label{fig1}
\end{figure}

For leptogenesis, the one-loop diagrams (for triplet fermion decays), Fig.~2, 
are exactly the same as for the right-handed neutrinos, Fig.~1, which lead to the same asymmetries up
to $SU(2)_L$ factors of order unity. The CP asymmetry ($M_{\Sigma_i}<<M_{\Sigma_j}$) is given as:
\begin{equation}\label{epsT}
\varepsilon_{\Sigma_i}=\sum_{j}\frac{3}{2}
  \frac{M_{\Sigma_i} }{M_{\Sigma_j} }\frac{\Gamma_{\Sigma_j} }{M_{\Sigma_j} }
  I_j\frac{V_j-2 S_j }{3} \,,
\end{equation}
As we know in hierarchical limit $S_j,V_j$ are of O(1). So the CP asymmetry from triplet decay
is suppressed by a factor of 3 in compare to $N_R$ decay. But the final amount of baryon asymmetry 
is given by CP asymmetry multiplied by an efficiency factor ($\eta$) and a numerical factor 
which is 3 times larger than that for singlet fermion decay. This is because the triplet 
fermions have three components. As pointed out earlier both have same decay width but the thermally 
averaged decay width is 3 times larger for triplet fermions again because of three components. 
The CP asymmetry from the self-energy and vertex correction diagrams of $\Sigma$ 
(for $M_{\Sigma_2} << M_{\Sigma_{1}}$):
\begin{equation}
\varepsilon_{\Sigma_2}=-\frac{1}{16 \pi} \frac{M_{\Sigma_2}}{M_{\Sigma_1}}
\frac {Im[(Y_{\Sigma} Y_{\Sigma}^{\dagger})_{21}^2]}
{\sum_i |(Y_{\Sigma})_{2i}|^2} \,,
\label{nu-nu}
\end{equation}
\begin{figure}[htb]
\centering \vspace*{3mm} \hspace*{1mm}
\epsfig{figure=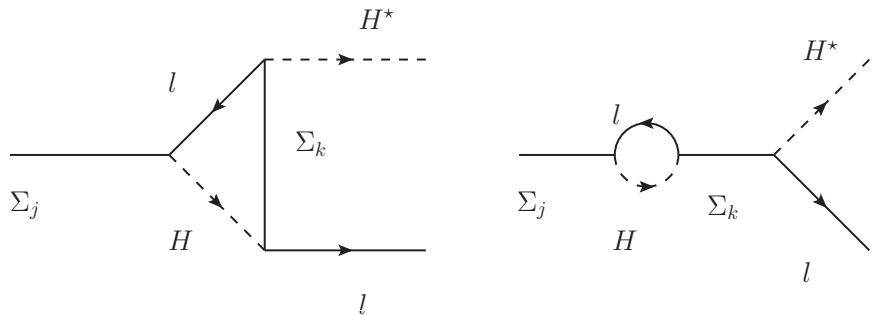,width =0.7\columnwidth} \vspace*{-1mm}
\caption{One loop vertex and self-energy diagrams contributing to
the asymmetry via decay of fermion triplet, $M_{\Sigma_j} <<
M_{\Sigma_{k}}$.} 
\label{fig3}
\end{figure}
We have hierarchical heavy fermions in our model: $M_{\Sigma_2}<M_{N_2}<M_{\Sigma_1}<M_{N_3}$. 
This allows us to have some more diagrams in vertex correction leading to CP asymmetry.
As $M_{N_2} < M_{\Sigma_{1}}$, the CP asymmetry from $N_2$ decay (with $\Sigma_1$ is in loop)
\begin{equation}
\varepsilon_{N_2}^{\Sigma_1}=-\frac{3}{16 \pi} \frac{M_{N_2}}{M_{\Sigma_1}}
\frac{\sum_{il}{\cal I}m[(Y_{N}^*)_{i2} (Y_{N}^*)_{l2}( Y_{\Sigma}^{\dagger} Y_{\Sigma})_{il}]}
{\sum_i |(Y_{N})_{2i}|^2} \,.
\label{nu-Sigma}
\end{equation}
Again there will be an extra contribution to 
CP asymmetry from decay of $\Sigma_2$ where $N_{2,3}$ will be in loop of the vertex diagram.
The contribution to CP asymmetry is:
\begin{equation}
\varepsilon_{\Sigma_2}^{N_{j}}=\frac{1}{16 \pi}\sum_j \frac{M_{\Sigma_2}}{M_{N_j}}
\frac{\sum_{il}{\cal I}m[(Y_{\Sigma}^*)_{i2} (Y_{\Sigma}^*)_{l2}( Y_{N}^{\dagger}Y_{N})_{il}]}
{\sum_i |(Y_{\Sigma})_{2i}|^2} \,.
\label{nu-Sigma}
\end{equation}

\section{Conclusions}
A variant $double$ seesaw mechanism for neutrino mass generation is discussed in the context of 
left-right symmetric theories where parity is spontaneously broken. Neutrino masses are generated 
by integrating out triplet and singlet fermions. In our model we have both heavy and TeV scale fermions. 
The decays of heavy right handed neutrinos and triplet fermions lead to leptogenesis. Their scale is not suitable for the LHC study of this model. But the triplet fermions and the right handed neutrinos having masses few TeV are within the reach of the LHC and ILC. Their possible signatures are discussed in the context of collider physics. The distinguishable final states: (i) six and five lepton final states, 
(ii) four leptons with total electric charge +2 - 
from triplet fermion decays along with the right handed neutrino signals make this model much more testable. 
In this model we can capture both leptogenesis and collider aspects as the heavy fermions coexist with the TeV scale fermions. This makes our model phenomenologically rich and testable. A detailed comparative study of the signals will be discussed in future publication in the context of the LHC and ILC \cite{CG}.

\vskip 20pt

{\bf Acknowledgments:} I would like to thank Biswarup Mukhopadhyaya and Amitava Raychaudhuri for many useful suggestions. This research has been supported by funds from the XIth Plan `Neutrino physics' and 
RECAPP projects at HRI.

\end{document}